\documentstyle [11pt]{article}

\renewcommand{\theequation}{\arabic{equation}}
\renewcommand{\thesubsection}{\thesection.\arabic{subsection}}

\def\tr{{\rm Tr}}

\def\overa#1{\stackrel{\! \! \! \leftrightarrow}{#1}}

\newcommand{\be}{\begin{eqnarray}}
\newcommand{\ee}{\end{eqnarray}}
\newcommand{\BE} {\begin{equation}}
\newcommand{\EE} {\end{equation}}
\newcommand{\BEAnn}{\begin{eqnarray*}}
\newcommand{\EEAnn}{\end{eqnarray*}}
\def\lrarr#1{\vbox{\ialign{##\crcr
    $\leftrightarrow$\crcr
    \noalign{\kern 1pt\nointerlineskip}
    $\hfil\displaystyle{#1}\hfil$\crcr}}}

\begin{document}

\thispagestyle{empty}
\hspace{4.4in}
OSU-NT-\#93-0122
\vspace{0.4in}

\setlength{\unitlength}{1in}
\begin{center}
\baselineskip 0.33in
{\LARGE\bf Light-Front QCD:} \\
{\Large\bf Role of Longitudinal Boundary Integrals}
\end{center}
\baselineskip 0.20in
\begin{center}
\  \\
\ \\
{\bf Wei-Min Zhang and Avaroth Harindranath} \\
\ \\
{\normalsize{\em Department of Physics, The Ohio State University \\
Columbus, Ohio 43210, USA}} \\
\end{center}
\vspace{0.4in}

\begin{abstract}
In the canonical light-front QCD, the elimination of unphysical
gauge degrees of freedom leads to a set of boundary integrals
which are associated with the light-front infrared singularity.
We find that a consistent treatment of the boundary integrals
leads to the cancellation of the light-front linear infrared
divergences. For physical states, the requirement of finite
energy density in the light-front gauge $(A_a^+=0)$ results in
equations which determine the asymptotic behavior of the transverse
(physical) gauge degrees of freedom at longitudinal infinity.
These asymptotic fields are generated by the boundary integrals
and they are responsible for the topological winding number.
They also involve non-local behavior in the transverse direction
that leads to non-local forces.
\end{abstract}
\vspace{0.5in}

\newpage

\section{Introduction}
Quantum chromodynamics (QCD) was initially proposed as a strong
interaction field theory {\em in} light-front coordinates,
motivated by light-front current algebra {\cite{gallmannqcd}}.
In recent years, the search for nonperturbative solutions of
QCD has led to an extensive exploration
of light-front field theory (LFFT).  The main attractions for
studying nonperturbative QCD in light-front coordinates,
called light-front form by Dirac {\cite{diracl}}, are that
{\cite{lflist}}: (1) boost invariance in LFFT is a kinematical
symmetry, which is important in the study of composite systems,
particularly the hadrons in QCD; (2) LFFT is a relativistic field
theory with nonrelativistic structure so that the relativistic
bound state equations are reduced to Schrodinger-type equations
from which the nonrelativistic quark model may find its
justification in QCD in light-front form; (3) the positivity
of the longitudinal momentum ($k^+ \geq 0$) in light-front Hamiltonian
field theory implies that the light-front vacuum consists only of
particles with longitudinal momentum $k^+=0$, which may simplify
the QCD vacuum structure.  These properties provide a hope to solve
QCD in light-front form for hadrons.

The earliest systematic formulation of the light-front QCD (QCD in
light-front form with light-front gauge, referred to simply as
LFQCD hereafter) was given about sixteen years ago {\cite{lfqcd1}}.
For applications of perturbative LFQCD see ref.{\cite{plfqcd}}.
In order to understand basic nonperturbative relativistic
bound state problems, in the last few years
many works on LFFT have mainly focused on various $1+1$
field theory models, and some on the 3+1 Yukawa model and
QED {\cite{nplfft,bmpp}}.  One main obstacle in extending the
study to nonperturbative LFQCD is that a formalism
to address simultaneously the major difficulties of QCD
in light-front form is still not in place.  These
difficulties include the renormalization problem
(even in perturbation theory), the confinement problem,
and the problem of the QCD vacuum and dynamical
chiral symmetry breaking.

The renormalization problem in the study of relativistic
bound states in LFFT has several aspects.  Since power
counting is different on the light-front {\cite{wilsonp}},
there are additional ultraviolet divergences in LFFT,
compared to the instant form {\cite{diracl}}. The additional
ultraviolet divergences have received some attention
recently in the context of the relativistic bound state problem in the
3+1 light-front Yukawa model {\cite{hari}}.  LFQCD also contains
severe light-front infrared divergences.  The resolution of the
light-front infrared divergence problems is not complete even in
perturbative LFQCD.  Issues arising from the possible mixing of
the ultraviolet and infrared divergences in the relativistic bound
state problems in LFFT have not been addressed so far.

Understanding confinement is crucial for building hadronic
bound states in QCD. In the present canonical LFQCD, the associated
Hamiltonian contains a linear potential between color charges
only in the longitudinal direction, which does not provide a
confinement mechanism for quarks and gluons in $3+1$ dimensions.
Therefore, it may not be suitable for describing low-energy
hadronic structure.  Based on light-front power
counting, Wilson recently proposed a formalism to construct
a confining light-front quark-gluon Hamiltonian for LFQCD
{\cite{wilsonp}}.  Wilson suggested that a starting point for
analyzing the full QCD with confinement in light-front form is
the linear infrared divergence (i.e., $1/k^{+2}$ singularity in
momentum space). The counterterms for the linear divergence,
which may be constructed from light-front power counting rules,
can involve the color charge densities and involve unknown
non-local behavior in transverse direction.  It is tempting
to identify these terms as the source of transverse confinement.
However, the analysis is not yet complete and
a scheme for practical calculation has yet to be developed.

Dynamical chiral symmetry breaking is another important
issue in the study of QCD for hadrons. In instant form,
dynamical chiral symmetry breaking is associated with a
nontrivial vacuum through the Goldstone
mechanism.  In LFQCD, the vacuum is trivial when the
$k^+=0$ sector is ignored.  Therefore, it seems to be natural
to argue that in order to obtain a nontrivial vacuum, one has
to solve the $k^+=0$ modes {\cite{bmpp}}.  Solving the $k^+=0$
modes may provide us with mechanism for spontaneous chiral
symmetry breaking.  Yet, the $k^+=0$ sector is singular and is
very ambiguous.  This singularity may exist even in free field
theory.  Thus, it is not clear whether the nontrivial structure of
LFQCD {\em must} be associated with the $k^+=0$ modes.
Furthermore, by involving the $k^+=0$ sector, the main
advantage of LFQCD that simplifies nonperturbative bound
states is lost, and therefore there is no strong reason
why we should study nonperturbative QCD in light-front form.
In fact, dynamical symmetry breaking can be manifested in
different ways in different frames.  It may be more attractive
if we could formulate LFQCD with a trivial vacuum such that the
dynamical breaking of chiral symmetry is manifested explicitly
via effective interactions.  However, the attempt in this
direction has not yet started.

All these problems mentioned above are essential and should
involve non-abelian gauge degrees of freedom in QCD.  We
are still unable to solve QCD at the moment.
As a starting point, we shall address in this
paper the problems of {\em light-front linear infrared
divergences and the associated nontrivial aspects}, based on
a canonical quantization approach to LFQCD.  We hope
that these discussions will provide some insight for
solving QCD in light-front form in the future.

We apply the conventional
canonical procedure{\cite{canonicalft}} to QCD in light-front
form. It turns out naturally that QCD is a generalized
Hamiltonian system {\cite{diracg}} where the first-class
gauge and quark constraints emerge explicitly in the
Lagrangian.  As is known, in the light-front gauge
these first-class constraints become solvable
first-order differential equations, and are used to eliminate
unphysical degrees of freedom to all orders of the coupling
constant.  However, the gauge constraint equations
contains a set of boundary integrals at longitudinal
infinity for the longitudinal color electric fields [see
eq.(15) and the following discussions].  These longitudinal boundary
integrals are the color charge density integrated over
the longitudinal space ($x^-$) and are associated with the
light-front infrared singularity.  The resulting LFQCD
Hamiltonian contains a boundary term proportional to these
boundary integrals which are overlooked in previous
investigations of light-front gauge theory.

We find that in perturbation theory the boundary integrals
serve to remove linear infrared divergences in loop integrals.
Removing the linear infrared divergences in LFQCD is a serious
problem that has not been solved completely.  In usual Feynman
theory of perturbative LFQCD, by use of the gauge fixing term,
one can derive the gauge propagator involving $1/k^+$
singularity.  Beyond the leading order calculation, this
singularity leads to linear infrared divergence in the principal
value prescription.  In $x^+$-ordering Hamiltonian
perturbation theory, the linear infrared divergences
emerge even in tree-level and one-loop diagrams.  By including
the boundary term in the Hamiltonian, we obtain a consistent
distribution function for the product of two principal value
prescriptions, from which the linear divergences in loop integrals
are removed by the same divergences in the instantaneous interaction.
This finding is useful for perturbative LFQCD calculations in
high-energy processes.

The relevant boundary integrals in the Hamiltonian formulation
of axial gauge were indeed pointed out first by Schwinger in 1962
{\cite{axialg}}.  Due to the different structure between
LFFT and the field theory in instant form, the consequences
from the boundary integrals we study in this
paper have not yet been realized explicitly in axial gauge.
One of the important differences is related to the QCD vacuum.
In instant form with axial gauge, the QCD vacuum cannot be
simple.  In LFQCD, generally the vacuum should also be nontrivial
because of the $k^+=0$ modes.  However, the choice of
antisymmetric boundary conditions for field variables at
longitudinal infinity excludes the $k^+=0$ modes.  In
this case, the LFQCD vacuum remains trivial as the bare
vacuum, and thus {\em the nontrivial QCD structure must
be carried purely by the boundary behavior of gauge fields.}

For physical states, the requirement of finite energy density
results in asymptotic equations for transverse (physical)
gauge fields at longitudinal infinity.
These asymptotic gauge fields are generated by the boundary
integrals.  It is these asymptotic fields that determine the
gauge field configurations for the non-vanishing winding number
associated with the topological solutions of QCD.
Therefore, although the LFQCD vacuum is trivial with our choice of
the boundary conditions, the nontrivial behavior of
gauge theory is manifested in field operators. To the best of our
knowledge, this is the first time the nontrivial behavior of
gauge fields is explored in light-front form with light-front
gauge $A_a^+=0$.  It is also the first attempt to address
the nontrivial structure with trivial vacuum in QCD,
which seems to be possible only in light-front form.

Furthermore, the asymptotic gauge fields at longitudinal infinity
which are generated by the boundary integrals not only involve
the color charge densities in transverse
space but also involve non-local behavior in the transverse
direction.  We find that by replacing the nontrivial boundary
condition with a trivial one for the transverse gluon fields,
many transverse non-local potentials are induced by the boundary
integrals.  These potentials are responsible for the nontrivial
QCD behavior and therefore may lead to quark and gluon confinement.
This possibility will be explored in further investigations.

The paper is organized as follows.  In section 2, a canonical
procedure for LFQCD is studied where we focus on the problem
of the boundary conditions in solving the light-front constraints.
In section 3, the roles of boundary integrals are explored in
detail.  Some remarks are made for relevant problems in section
4.  Finally, the paper includes two appendices.  In appendix A,
we discuss canonical quantization of LFQCD by use of the rigorous
phase space structure {\cite{fjc}} rather than the Dirac procedure
{\cite{diracq}}.  In appendix B, we demonstrate
the cancellation of linear infrared divergences at the tree level
in $q\bar{q}$ scattering and in one-loop diagrams for the quark mass
correction.

\section{Canonical formulation and boundary condition}
We start from the QCD Lagrangian
\BE
	{\cal L} = - \frac{1}{2} \tr (F^{\mu \nu}F_{\mu \nu}) + \bar{\psi}
		( i \gamma_{\mu} D^{\mu} - m ) \psi ,
\EE
where $F^{\mu \nu} = \partial^{\mu} A^{\nu} - \partial^{\nu} A^{\mu} -
i g [A^{\mu},  A^{\nu} ]$, $A^{\mu} = \sum_a A_a^{\mu} T^a$ is a
$3 \times 3$ gluon field color matrix and the $T^a$ are the generators
of the $SU(3)$ color group: $[T^a, T^b] = if^{abc}T^c$ and $\tr
(T^a T^b) = \frac{1}{2}\delta_{ab}$. The field variable $\psi$
describes quarks with three colors and $N_f$ flavors, $D^{\mu}
= \frac{1}{2} \overa{\partial^{\mu}} - i g A^{\mu} $ is the
symmetric covariant derivative, and $m$ is an $N_f \times N_f$
diagonal quark mass matrix.  The Lagrange equations of motion are
\be
	& & \partial_{\mu} F_a^{\mu \nu} + g f^{abc} A_{b \mu} F_c^{\mu \nu}
		+ g \bar{\psi} \gamma^{\nu} T^a \psi =0, \\
	& & (i \gamma_{\mu} \partial^{\mu} - m + g \gamma_{\mu} A^{\mu})
		\psi = 0.
\ee

The light-front coordinates are defined as: $x^{\pm} \equiv x^0
\pm x^3 ,~x_{\bot}^i \equiv x^i (i=1,2)$, where $x^+$ is chosen
as the ``time'' direction along which the states are evolved,
and $x^-$ and $x_{\bot}$ become naturally
the longitudinal and transverse coordinates.  The inner product
of any two four-vectors is then $a_{\mu} b^{\mu} = \frac{1}{2}(a^+
b^- + a^- b^+) - a_{\bot} \cdot b_{\bot}$, and the time and space
derivatives $(\partial^{\mu} = \frac{\partial}{\partial x_{\mu}})$
and the 4-dimensional volume element are given by $\partial^+ = 2
\frac{\partial}{\partial x^-}$, $\partial^- = 2 \frac{\partial}{
\partial x^+}$, $\partial^i = - \frac{\partial}{\partial x^i}$,
and $d^4x = \frac{1}{2} dx^+ dx^- d^2 x_{\bot}$, respectively.

Naively, the canonical theory of QCD in light-front form
is constructed by defining the conjugate momenta of field
variables $\{ A_a^{\mu}(x), \psi(x), \overline{\psi}(x) \}$
as
\be
	& & E_a^{\mu}(x) = \frac{\partial{\cal L}}{\partial
		(\partial^- A_{a \mu})} = - \frac{1}{2} F_{a}^{+ \mu}(x), \\
	& & \pi_{\psi} (x) =  \frac{\partial{\cal L}}{\partial
		(\partial^- \psi)} = i \frac{1}{4} \bar{\psi} \gamma^+
		= \frac{i}{2} \psi_+^{\dagger}(x) , \\
	& & \pi_{\psi^{\dagger}} (x) =  \frac{\partial{\cal L}}{\partial
		(\partial^- \psi^{\dagger})} = - i \frac{1}{4} \gamma^0
		\gamma^+ \psi = - \frac{i}{2} \psi_+(x) ,
\ee
where the fermion spinor in light-front form is divided into
$\psi = \psi_+ + \psi_-$, $\psi_{\pm} = \Lambda_{\pm} \psi$ with
$ \Lambda_{\pm} \equiv \frac{1}{2} \gamma^0 \gamma^{\pm} $.
Following a similar procedure in instant form described by Faddeev
and Slavnov{\cite{canonicalft}} for gauge theory, we separate the
time derivative terms from the Lagrangian,
\be
	& & {\cal L} = \left\{ \frac{1}{2} F_a^{+i} (\partial^- A_a^i) +
		\frac{i}{2} \psi_+^{\dagger} (\partial^- \psi_+) - \frac{i}{2}
		(\partial^- \psi_+^{\dagger}) \psi_+ \right\} - {\cal H}
			\nonumber \\
	& & ~~~~~~~~~~~~~~~~~~- \left\{ A_a^- {\cal C}_a + \frac{1}{2}
		(\psi_-^{\dagger} {\cal C} + {\cal C}^{\dagger} \psi_- )
		\right\} ,
\ee
where
\be
	& & {\cal H} = \frac{1}{2} ( E_a^{-2} + B_a^{-2} ) + \frac{1}{2}
		\left\{ \psi_+^{\dagger} \{ \alpha_{\bot} \cdot (i\partial_{\bot}
		+ g A_{\bot}) + \beta m\} \psi_- + h.c. \right\} \nonumber \\
	& & ~~~~~~~~~~~~~~~~~~~+ \left\{ \frac{1}{2} \partial^+(E_a^- A_a^-)
		- \partial^i (E_a^i A_a^-) \right\}
\ee
and
\be
	& & {\cal C}_{a} = \frac{1}{2}(\partial^+ E_a^- + g f^{abc} A_b^+
		E_c^-) - (\partial^i E_a^i + g f^{abc} A_b^i E_c^i)
		+ g \psi_+^{\dagger} T^a \psi_+ \\
	& & {\cal C} = (i \partial^+ + g A^+) \psi_-
		- (i \alpha_{\bot} \cdot \partial_{\bot} + g
		\alpha_{\bot} \cdot A_{\bot} + \beta m) \psi_+ .
\ee
In eq.(8), we have defined $B_a^- = F_a^{12}$ as the longitudinal component
of the light-front color magnetic field.

The reason for writing the Lagrangian in the above form is to make
the Hamiltonian density and also the dynamical variables and
constraints manifest.  In eq.(7), the first term contains
all the light-front time derivative terms.  From the
definition of eqs.(4-6), it immediately follows that only
the transverse gauge fields $A_a^i$ and the up-component
quark fields $\psi_+$ and $\psi_+^{\dagger}$ are dynamical
variables.  The second term in eq.(7), ${\cal H}$, is a Hamiltonian
density.  It contains three parts, the first part involves
the light-front color electric and magnetic fields; the
second, the usual quark Hamiltonian
with coupling to the gauge field, and the last a surface term.
Besides the kinetic term and the Hamiltonian density, eq.(7)
also contains an additional term.  This is a constraint term
which indicates that the longitudinal gauge
field $A_a^-$ and the down-component quark fields $\psi_- ~
(\psi_-^{\dagger})$ are only the Lagrange multipliers for the
constraints ${\cal C}_{a},{\cal C}~({\cal C}^{\dagger})=0$.
These constraints arise from the definition of canonical
momenta in the light-front coordinates and are consistent
with the Lagrangian equations of motion.  The gauge field
constraint, ${\cal C}_a = 0$, is in fact the light-front
Gauss law which is an intrinsic property of gauge theory.
The fermion constraint, ${\cal C}~({\cal C}^{\dagger}) = 0$ is
purely a consequence of using the light-front form.

The existence of constraint terms simply implies that QCD in the
light-front form is a generalized Hamiltonian system{\cite{diracg}}.
These constraints are all secondary, first-class
constraints{\footnote{In light-front field theory, there always
exist the so-called primary, second-class constraints due to
the fact that the light-front Lagrangian is linear in the first-order
$x^+$-derivative and therefore the canonical momenta are
functions of field variables.  These constraints are not real
constraints in the generalized phase space quantization
procedure and are easily handled.  We will put these discussions
in the Appendix A.}} in the Dirac procedure of quantization
{\cite{diracq}}.  To obtain a canonical formulation of
LFQCD for non-perturbative calculations, we need to explicitly
solve the constraints, namely to determine the Lagrange
multipliers, to all orders of the coupling constant.  Generally, it
is very difficult to analytically determine the Lagrange
multipliers from the constraints ${\cal C}_a, {\cal C}=0$
since they are coupled by $A_a^+$.  Only
in the light-front gauge {\cite{lfqed,lfqed1}},
\BE
	A_a^+ (x) \equiv A_a^0(x) + A_a^3(x) = 0,
\EE
are these two constraints reduced to solvable one-dimensional
differential equations:
\BE
	\left\{ \begin{array}{l}
		\frac{1}{2} \partial^+ E_a^- = \partial^i E_a^i + g (f^{abc}
		A_b^iE_c^i - \psi_+^{\dagger} T^a \psi_+) \equiv G_a \\ ~~ \\
		i \partial^+ \psi_- = ( i \alpha_{\bot} \cdot \partial_{\bot}
		+ g \alpha_{\bot} \cdot A_{\bot} + \beta m) \psi_+
		\end{array} \right. .
\EE

In order to solve eq.(12), we have to define the operator $1/\partial^+$.
In general,
\BE
	\left(\frac{1}{\partial^+}\right) f(x^-,x^+,x_{\bot}) =
		\int_{-\infty}^{\infty} dx_1^- \varepsilon(x^- - x_1^-)
		f(x_1^-,x^+,x_{\bot}) + C(x^+,x_{\bot})
\EE
where $\varepsilon(x) =1, 0 , -1 $ for $x > 0, =0, <0$, respectively,
and $C(x^+,x_{\bot})$ is a $x^-$ independent constant. However,
since the canonical conjugate of transverse
gauge field in LFQCD is a dependent variable [$E_a^i = - \frac{1}{2}
\partial^+ A_a^i $, see eq.(4) with $A_a^+=0$], one has to impose {\em a
priori} a boundary condition for $A_a^i$ in order to derive the
canonical commutation relations for the physical field variables. It has
been shown {\cite{lfqed1,jackiw2}} that the suitable definition
of $1/\partial^+$ which uniquely determines the initial value
problem at $x^+=0$ for independent field variables is $C(x^+,
x_{\bot}) = 0$.  This corresponds to choosing an antisymmetric
boundary condition for field variables in the longitudinal direction.

Using eq.(13), we can explicitly express $E_a^-$ in terms of
transverse gauge fields $A_a^i$ and the independent light-front
quark field $\psi_+$ from the gauge constraint in eq.(12),
\BE
	E_a^- (x) +  \partial^i A_a^i (x) = - \frac{g}{4} \int_{-\infty}^{\infty}
		dx'^- \varepsilon (x^- - x'^-) (f^{abc} A_b^i \partial^+ A_c^i
		+ 2 \psi_+^{\dagger} T^a \psi_+) + C_a(x^+,x_{\bot}),
\EE
here, $E_a^i = - \frac{1}{2} \partial^+ A_a^i$ has been used.  To
uniquely determine their initial values at $x^+=0$, we require
that the $E_a^-$ and $A_a^i$ satisfy antisymmetric boundary
conditions at longitudinal infinity, namely, $C_a(x^+,x_{\bot}) = 0$.
As a result,
\be
	& & E_a^-(x) = -\partial^i A_a^i (x) - \frac{g}{4}
		\int_{-\infty}^{\infty} dx'^- \varepsilon (x^- - x'^-)
		(f^{abc} A_b^i \partial^+ A_c^i + 2 \psi_+^{\dagger} T^a
		\psi_+) \nonumber \\
	& & ~~~~~~ \stackrel{x^- = \pm \infty}{=}
		-\partial^i A_a^i|_{x^-=\pm \infty}  \mp
		\frac{g}{4} \int_{-\infty}^{\infty} dx^- (f^{abc} A_b^i
		\partial^+ A_c^i + 2 \psi_+^{\dagger} T^a \psi_+).
\ee

Since $E_a^-$ satisfies now an antisymmetric boundary condition, its
boundary values at longitudinal infinity are completely determined
by the second equality in eq.(15), where the second term is boundary
integrals over $x^-$ for the color charge densities.  These integrals
are the source of light-front infrared singularity.  {\em We call
them the longitudinal boundary integrals,
or simply the boundary integrals}.

By using the identity{\cite{axialg}},
\BE
	\frac{1}{2} \int_{-\lambda/2}^{\lambda/2} dx^- \varepsilon(x^- - x'^-)
		\varepsilon(x^- - x''^-) = - | x'^- - x''^- | + \frac{1}{2}
		\lambda ,
\EE
where the parameter $\lambda$ denotes the distance between two boundary points
in the longitudinal direction, the color electric field energy in the
Hamiltonian becomes
\be
	& & H_E = \frac{1}{2} \int_{-\infty}^{\infty} dx^- d^2x_{\bot}
		(E_a^-)^2 = \frac{1}{2} \int_{-\infty}^{\infty} dx^-
		d^2x_{\bot} \left\{ \frac{ }{ }(\partial^iA_a^i)^2
		\right. \nonumber \\
	& & ~~~~~~~~~~~~~~~~~~~~~ + \frac{1}{2} \int_{-\infty}^{\infty} dx'^-
		\left[ g \partial^i A_a^i \varepsilon(x^- - x'^-) (f^{abc}
		A_b^i \partial^+ A_c^i + 2 \psi_+^{\dagger} T^a \psi_+ )
		\right. \nonumber \\
	& & ~~~~~~~~~~~~~~~~~~~~~ - \frac{g^2}{8} \int_{-\infty}^{\infty}
		dx'^- (f^{abc} A_b^i \partial^+ A_c^i + 2 \psi_+^{\dagger}
		T^a \psi_+ ) \nonumber \\
	& & ~~~~~~~~~~~~~~~~~~~~~~~~~~~~~~~~~~~ \left. |x^- - x'^-| (f^{ade}
		A_d^j \partial^+ A_e^j + 2 \psi_+^{\dagger} T^a \psi_+ )
		\frac{ }{ } \right\} \nonumber \\
	& & ~~~~~~~~~~~~~~~~~~~~~ + \left( \lim_{\lambda \rightarrow \infty}
		\lambda \right) \frac{g^2}{8} \int d^2 x_{\bot} \left\{ \int_{-
		\infty}^{\infty}dx^- (f^{abc} A_b^i \partial^+ A_c^i +
		2 \psi_+^{\dagger} T^a \psi_+) \right\}^2.
\ee
In the above equation, the last term (a boundary term) involves the
boundary integrals and is associated with the infrared
divergence in the light-front instantaneous interactions.
As we will discuss later, in perturbation theory, this term is
regularized by the distribution function of the product of two
principal value prescriptions and leads to the cancellation of
the light-front linear infrared divergences.  For physical states,
the requirement of finite energy density results in the asymptotic
equations for the transverse gauge fields which show that the
asymptotic transverse gauge fields do not vanish at
longitudinal infinity and are generated by the boundary
integrals.  The non-vanishing asymptotic transverse gauge fields
determine the nontrivial QCD structure in light-front form. Thus,
the boundary integrals can inherently affect QCD dynamics.

The Lagrange multipliers in eq.(7) can be determined easily now.
The Lagrange multiplier $\psi_-$ is the solution of the quark
constraint in eq.(12),
\be
	& & \psi_- (x) = - \frac{i}{4} \int_{-\infty}^{\infty}
		dx'^-d^2x_{\bot}' \varepsilon (x^- - x'^-)
		\delta^2 (x_{\bot} - x'_{\bot}) \nonumber \\
	& & ~~~~~~~~~~~~~~~~~~~~~~~ \times [\alpha_{\bot} \cdot ( i
		\partial'_{\bot} + g A_{\bot}(x')) + \beta m] \psi_+ (x').
\ee
The Lagrange multiplier $A_a^-$ is obtained from the definition $E_a^-
= - \frac{1}{2} \partial^+ A_a^-$ and eq.(15),
\BE
	A_a^- (x) = - \frac{1}{2} \int_{-\infty}^{\infty} dx'^-
		\varepsilon(x^- - x'^-) E_a^-(x^+,x'^-,x_{\bot})
\EE
For this solution, the first surface term in eq.(8) vanishes.
Moreover, it is reasonable to assume that the transverse color
electric fields $E_a^i$ as well as $A_a^i$ vanish as $O(r^{-2})$
and $O(r^{-1})$ at $r=|x_{\bot}|
\rightarrow \infty$ because the gauge freedom is totally fixed at
the transverse infinity.  Hence the other surface term in eq.(8)
vanishes as well.

After the determination of the Lagrange multipliers, the LFQCD
Hamiltonian is given simply by
\be
	& & H = \int dx^- d^2 x_{\bot} \left\{ \frac{1}{2} ( E_a^{-2}
		+ B_a^{-2}) + \psi_+^{\dagger} \{ \alpha_{\bot}^i
		( i \partial_{\bot}^i + g A^i) + \beta m \} \psi_-
		\right\}  \nonumber \\
	& & ~~~~ = \int dx^- d^2 x_{\bot} \left\{ \frac{1}{2} (\partial^i
		A_a^j)^2 + g f^{abc}A_a^i A_b^j \partial^i A_c^j
		+ \frac{g^2}{4} f^{abc}f^{ade} A_a^i A_b^j A_d^i A_e^j
		\right. \nonumber \\
	& & ~~~~~~~~~~ + \frac{1}{4} \int_{-\infty}^{\infty} dx'^-
		\left[ g \partial^i A_a^i \varepsilon(x^- - x'^-) (f^{abc}
		A_b^i \partial^+ A_c^i + 2 \psi_+^{\dagger} T^a \psi_+ )
		\right. \nonumber \\
	& & ~~~~~~~~~~~~~~~ \left. -i \psi_+^{\dagger} \{ \alpha_{\bot}^i
		( i \partial_{\bot}^i + g A^i) + \beta m \} \varepsilon(x^-
		- x'^-) \{ \alpha_{\bot}^j ( i \partial_{\bot}^j + g A^j)
		+ \beta m \} \psi_+ \right]  \nonumber \\
	& & ~~~~~~~~~~ - \frac{g^2}{16} \int_{-\infty}^{\infty} dx'^-
		(f^{abc} A_b^i \partial^+ A_c^i + 2 \psi_+^{\dagger}
		T^a \psi_+ ) \nonumber \\
	& & ~~~~~~~~~~~~~~~~~~~~~~~~~~ \left. |x^- - x'^-| (f^{ade} A_d^j
		\partial^+ A_e^j + 2 \psi_+^{\dagger} T^a \psi_+ )
		\right\} \nonumber \\
	& & ~~~~~~~~~~ + \left( \lim_{\lambda \rightarrow \infty} \lambda
		\right) \frac{g^2}{16} \int d^2 x_{\bot} \left\{ \int_{-
		\infty}^{\infty}dx^- (f^{abc} A_b^i \partial^+ A_c^i +
		2 \psi_+^{\dagger} T^a \psi_+) \right\}^2.
\ee
A detailed procedure to quantize the above formulation is presented
in appendix A.  This Hamiltonian contains the naive canonical LFQCD
Hamiltonian plus a boundary term which is the square of the boundary
integrals, as a result of eq.(15). Choosing antisymmetric
boundary conditions in the longitudinal direction in LFQCD
has the following advantages:

1). With any other boundary condition, eq.(14) contains
an arbitrary $x^-$-independent function.
Such an arbitrary function leads to ambiguities in formulating
LFQCD.  Only with an antisymmetric boundary conditions, is this
arbitrary term zero and formally LFQCD can be completely defined.

2). For the definition eq.(13) with $C(x^+,x_{\bot})=0$, all field
variables in LFQCD satisfy antisymmetric boundary conditions
at longitudinal infinity except $\psi_+(x)$, whose boundary
condition is not specified.  However, the equation of motion for
$\psi_+ (x)$ contains $(\frac{1}{\partial^+}) \psi_+$ [see Appendix
A] which forces it to satisfy the antisymmetric boundary condition.
As a result, the $k^+=0$ modes are completely excluded
in the momentum expansion of field variables.  Since the LFFT
vacuum is occupied only by the $k^+=0$ particles, with the help of
antisymmetric boundary condition the LFQCD vacuum is ensured
to be trivial as the bare vacuum.  The important consequence
is, as we shall discuss in the next section, that in this case
the nontrivial structure of QCD is carried purely by field operators.

3). By the choice of antisymmetric boundary conditions, the residual
gauge freedom in $A_a^+=0$ is completely fixed {\cite{zhang3}}
(also see the discussion later).

It is worth pointing out that there is a disadvantage for using
eq.(13) in perturbative light-front Feynman loop integrals. In perturbative
theory, eq.(13) with $C=0$ leads to the principal value prescription
for the light-front longitudinal infrared singularity.  Although
this prescription removes severe linear infrared divergences
as we will show next, severe logarithmic infrared divergences
are still present.  These logarithmic infrared singularities correspond
to the ``spurious'' poles in Feynman integrals, which prohibit any
continuation to euclidean space (Wick rotation) and hence the
use of standard power counting arguments for Feynman loop
integrals {\cite{Caper}}.  This causes difficulties in addressing
renormalization of QCD in Feynman perturbation theory with light-front
gauge. In the last decade there are many investigations in attempting
to solve this problem.  One excellent solution is given by Mandelstam
and Leibbrandt, i.e., Mandelstam-Leibbrandt (ML) prescription {\cite{MLp}},
which allow continuation to euclidean space and hence power counting.
This prescription has also been derived in the equal-time quantization
with light-front gauge by Bassetto et al. {\cite{bassetto}}.
It has been also shown that with ML prescription, the {\em multiplicative}
renormalization in the two-component LFQCD Feynman formulation is
restored {\cite{lee}}.

In the present paper, we study QCD in light-front equal-$x^+$
quantization. Unfortunately, ML prescription cannot be applied
to equal-$x^+$ quantization because ML prescription is
defined on the boundary condition which involves $x^+$ itself
{\cite{Tang}} and are not allowed in equal-$x^+$ canonical theory.
Yet, as is pointed out recently by Wilson {\cite{wilsonp}},
light-front power counting differs completely from the power counting
in equal-time quantization.  Furthermore, the current attempts to
understand nonperturbative QCD in light-front form is based on the
$x^+$-ordered (old-fashioned) diagrams in which no Feynman
integral is involved {\cite{nplfft,bmpp}}. Thus the problem with
the power counting criterion in Feynman loop integrals does not
affect our discussions in this paper. The renormalization of light-front
Hamiltonian is an entirely new subject where investigations are still
in their preliminary stage {\cite{wilsonp,hari}}.

The main aim of this paper is to show that the boundary integrals
play a important role in understanding the nontrivial features of
LFQCD.  The logarithmic infrared divergences are completely
cancelled in the complete loop diagrams of dynamical processes,
as was previously shown in the calculation of QCD correction
to the scale evolution of hadronic structure function up to
two-loop {\cite{curci}}.  A simple example of such a
cancellation in $x^+$-ordered perturbative theory is also
given for quark mass renormalization in Appendix B-2.  In our
forthcoming papers {\cite{zhang1}} we will present a detailed
discussion on $x^+$-ordered perturbative loop calculations and
light-front renormalization in LFQCD Hamiltonian theory, where
the logarithmic infrared divergences are again completely
cancelled in coupling constant renormalization. However,
based on light-front power counting, the linear infrared
divergences only involve color charge density and involve non-local
behavior in the transverse direction, which may be the source of
transverse confinement in LFQCD {\cite{wilsonp}}.  The severe
linear infrared divergences in LFQCD has not been explored in
light-front Hamiltonian.  In the present paper, we shall focus
on linear infrared singularity associated with the boundary integrals,
which may relate to the nonperturbative aspects of LFQCD in
physical states, as we will see below.

\section{Role of boundary integrals}
1. {\em Removing linear infrared divergences}.
In the past decade, applications of LFQCD are mostly restricted
in perturbation theory.  Naively, the boundary term in eq.(20)
is ignored so that the light-front instantaneous interactions are
thought to be linear potentials {\cite{lfqed,yan}}.  However, this
negligence leads to severe infrared singularities in the perturbation
theory.  To see this
clearly, we consider the formulation in momentum space.  For the
prescription of $1/\partial^+$ expressed in terms of the integral
of eq.(13), the standard Fourier transform leads to the principal
value prescription in momentum space as follows,
\be
	& & \left(\frac{1}{\partial^+} \right) f(x^-) = \frac{1}{4}
		\int_{-\infty}^{\infty}dx'^- \varepsilon(x^- - x'^-)
		f(x'^-) \nonumber \\
	& & ~~~~~~~~~~~~~~~~ \longrightarrow
		\frac{1}{2} \left(\frac{1}{k^+ + i \epsilon}
		+ \frac{1}{k^+ - i \epsilon} \right) f(k^+)
		\equiv \frac{1}{[k^+]} f(k^+), \\
	& & \left(\frac{1}{\partial^+} \right)^2 f(x^-) = \frac{1}{4^2}
		\int_{-\infty}^{\infty}dx'^- dx''^- \varepsilon(x^- - x'^-)
		\varepsilon(x'^- - x''^-) f(x''^-) \nonumber \\
	& & ~~~~~~~~~~~~~~~~ \longrightarrow
		\left[\frac{1}{2} \left(\frac{1}{k^+ + i \epsilon}
		+ \frac{1}{k^+ - i \epsilon} \right) \right]^2 f(k^+)
		\equiv \frac{1}{[k^+]^2} f(k^+).
\ee
Eq.(22) defines the product of two principal value prescriptions of
eq.(21) in terms of the distribution function.  In this derivation,
it follows (see eq.(16)) that the boundary term in eq.(20) have
been regularized.  It is known that eq.(22) leads to linear infrared
divergences in loop integrals.  In order to avoid this divergence,
one naively introduces the following prescription {\cite{plfqcd}}
\be
	& & \left(\frac{1}{\partial^+} \right)^2 f(x^-) = \frac{1}{4^2}
		\int_{-\infty}^{\infty}dx'^- |x^- - x'^-|
		f(x'^-) \nonumber \\
	& & ~~~~~~~~~~~~~~~~ \longrightarrow
		\frac{1}{2} \left(\frac{1}{(k^+ + i \epsilon)^2}
		+ \frac{1}{(k^+ - i \epsilon)^2} \right) f(k^+)
		\equiv \frac{1}{[k^{+2}]} f(k^+).
\ee
This corresponds to the case that the longitudinal boundary
term in eq.(16) is ignored.  Equivalently, the last term
in eq.(20) is dropped.  Apparently, this prescription removes
the linear infrared divergence originated from the instantaneous
interactions.  Unfortunately in such a prescription, beyond leading
order calculations in Feynman perturbation theory or even
in leading order calculation in the old-fashion
Hamiltonian perturbation theory, the product
of two principal value prescriptions appearing from
three-point vertex either is not defined or leads
to linear infrared divergences.   We shall show that it
is the prescription of eq.(22) which serves for the cancellation
of linear infrared divergences originated from the three-point
vertex and from the instantaneous interactions.  Here we only discuss
the $x^+$-ordered (old-fashion) perturbative calculations.

First at the tree level, for example for $q \bar{q}$
scattering (see appendix B-1), only the linear potential leads to
a $1/\epsilon^2$ divergence as $k^+ \rightarrow 0$.  The scattering
involving one-gluon-exchange is finite due to the principal value
prescription.  Thus in the naive prescription (23), even the lowest
order $q\bar{q}$ scattering amplitude is $1/\epsilon^2$ divergent.
By including the boundary term, this divergence is cancelled.
In loop calculations, for example for the one-loop correction
to the quark self-energy (see appendix B-2), the one gluon
exchange diagram (which contains an integral of $1/[k^+]^2$)
leads to a $1/\epsilon$ divergence; the linear potential is,
however, infrared finite in the relevant integral of $1/[k^{+2}]$.
Hence, in the naive prescription, loop calculations
also contain the severe $1/\epsilon$ infrared divergence.
In prescription (22), the instantaneous interactions
are the linear potentials accompanied by the boundary term,
which produce a $1/\epsilon$ divergence
from the integral of $1/[k^+]^2$ that cancels precisely the
same divergence in the one gluon exchange diagram.  Furthermore,
the cancellation in the one-loop correction of quark-gluon vertex
has also been verified {\cite{zhang1}}.

The reason that the linear infrared divergences are removed by
using prescription (22) can be understood as follows.
{}From eq.(15), the $k^+$ singularity originated from the
boundary integrals.  The color electric energy in LFQCD Hamiltonian
contains two sources for the $k^+$ singularity.  One is
the explicit boundary term, the last term in eq.(20),
which is $1/k^{+2}$-singular.  The other belongs to the
gluon emission vertex.  The resultant gluon emission
vertex is the first term in the square bracket in
eq.(20), which is $1/k^+$-singular.  Therefore, in one gluon
exchange diagrams, it produces a $1/k^{+2}$-singularity,
namely the product of two principal value prescriptions for
the definition of eq.(21).  The associated linear divergence
in loop integrals is the same as that from the $1/k^{+2}$-singularity
of the boundary terms in the prescription of eq.(22),
with a different sign from an energy denominator, and therefore
the linear infrared divergence is cancelled.  Note that in
eq.(20) there is another $1/k^+$-singularity
[in the second term in the square bracket], which comes from the
quark constraint [see eq.(18)].  Yet, in one-gluon exchange
diagrams, it leads to a form $1/(p_1^+p_2^+)$ ($p_1^+=p_2^++k^+$)
which does not generate infrared divergences.  Thus, all linear
infrared divergences originate from the same source, the boundary
integrals. Any negligence of boundary term in the Hamiltonian
through eq.(23) will lead to unwanted infrared divergences.

However, the cancellation of the linear infrared divergences in
higher order loop-integrals (beyond the one-loop diagrams) may
also depend on the regularization of ultraviolet
divergences.  The cancellation beyond leading order should be true
for gauge invariant regularization.  For gauge variant regularization,
such as transverse dimensional regularization {\cite{lfqcd1}},
boost invariant cutoff regularization {\cite{imr}} and the explicit cutoff
regularization {\cite{ecr}} used in the $x^+$-ordered perturbative
LFQCD, we need to introduce gluon mass counterterms.  These
counterterms break gauge invariance
and thereby may also spoil the cancellation of linear infrared
divergences in higher order diagrams.
However, if we set the quark mass $m=0$ in perturbative LFQCD, the
transverse dimensional regularization results in a zero gluon
mass correction.  In this case the cancellation is still satisfied
in two-loop diagrams.  In deep inelastic scattering, one often sets
$m=0$ in calculating high-order corrections to the scale evolution
of hadronic structure functions {\cite{curci}}.
A more detailed discussion on perturbative LFQCD will be presented
in a separate paper {\cite{zhang1}}.  For low-energy dynamics, the light
quark mass is crucial and perturbation theory is no longer useful.
Removes infrared divergences needs to be treated in an alternative way,
which we shall discuss later.

We may point out that in $1+1$ LFQCD {\cite{thooft}}, the
boundary integral is the color charge operator.  The corresponding
boundary term occurring in the Hamiltonian {\cite{qcd111}} is
then proportional to the square of color charge.
It is indeed this term resulting in an infinite quark
mass which is regarded as evidence of quark
confinement in $1+1$ QCD.  Explicitly, the linear potential
does not provide an infinite mass for the quark,
as shown above (also see ref.{\cite{qcd112}}),
but the boundary term adds a $1/k^{+2}$ singularity
(the $1/\epsilon$ divergence) to the quark propagator.
Since there are no transverse gluons in $1+1$ QCD to
cancel this divergence, the boundary term recovers
't Hooft's solution of the infinite quark mass pole{\cite{thooft}}.
In physical (zero color charge) states, the boundary term does not
contribute to physical observables since it is the
square of the color charge operators.  Quark confinement
in gauge-invariant states arises purely from the linear potential.
This implies that ignoring the boundary integral in $1+1$ QCD
may not affect any observable.  In $3+1$ QCD, the existence of
transverse gluons changes these consequences.

2. {\em Winding number}. The winding number is a topological
quantity associated with nontrivial structure of U(1) problem in
QCD (in Euclidean space, this is the topological charge or Pontryagin
index associated with the instanton solution of non-abelian gauge
theory) {\cite{monopole,coleman}}. The second important consequence
from the boundary integrals is that they determine a non-vanishing
winding number.

As it is known $A_a^+=0$ cannot completely fix the gauge
degrees of freedom.  There are residual gauge transformations
under which the theory is invariant.  The generators of
residual gauge transformations are
\BE
	R_a = - \frac{1}{2} \int_{-\infty}^{\infty} dx^-
		\left\{ 2 \partial^i \partial^+ A_a^i
		+ g(f^{abc} A_b^i \partial^+ A_c^i + 2
		\psi_+^{\dagger} T^a \psi_+) \right\}
\EE
As we have shown {\cite{zhang3}} the residual gauge transformations
generated by $R_a$ break the antisymmetric boundary condition for
$A_a^i$ at longitudinal infinity and therefore are not allowed
with our choice of the boundary conditions.  However, by using the
antisymmetric boundary conditions, the first term
in eq.(24) can be integrated out explicitly over $x^-$ and we have,
\BE
	R'_a = \mp 4\partial^i A_a^i |_{x^- = \pm \infty}
		- \frac{g}{2} \int_{-\infty}^{\infty} dx^-
		(f^{abc} A_b^i \partial^+ A_c^i + 2
		\psi_+^{\dagger} T^a \psi_+).
\EE
It has been verified {\cite{zhang3}} that for this definition, $R'_a$
generate the gauge transformations preserving the antisymmetric
boundary conditions.  But these operators do not commute with the
LFQCD Hamiltonian (20).  In other words, there is no longer an additional
gauge freedom to choose other $A_a^i$ such that the resulting Hamiltonian
remains invariant.  Therefore, with antisymmetric boundary conditions,
the residual gauge freedom in $A_a^+=0$ is completely fixed {\cite{zhang3}}.

Furthermore, for physical states, finite energy density requires
that the longitudinal color electric field strength must vanish in
the infinity (A similar requirement was used by Chodos in axial gauge
{\cite{axialg}}):
\BE
	E_a^- |_{x^- = \pm \infty} = 0
\EE
or explicitly
\BE
	\partial^i A_a^i |_{x^- = \pm \infty} =
		\mp \frac{g}{4} \int_{-\infty}^{\infty} dx^- (f^{abc}
		A_b^i \partial^+ A_c^i + 2 \psi_+^{\dagger} T^a \psi_+) .
\EE
Eq.(27) is consistent with our choice of antisymmetric boundary
condition.  Moreover, this condition explicitly shows that the
transverse gauge fields at longitudinal infinity are generated
by the boundary integrals.{\footnote{It may be worth pointing
out that unlike the $1+1$ QCD, the boundary integrals in 3+1
LFQCD are color charge densities in the transverse space and
not the color charge operators. The color charge operator
is defined as follows:
\[ \begin{array}{l}
	Q_a =  \int dx^- d^2x_{\bot} (f^{abc} A_b^i \partial^+
		A_c^i + 2 \psi_+^{\dagger} T^a \psi_+) \\
	~~~~~= \int dx^- d^2x_{\bot} (\rho_a^g(x^-,x_{\bot})
		+ \rho_a^q(x^-,x_{\bot})) \end{array} \]
where $\rho_a^g$ is the charge density carried by gluon field
and $\rho_a^q$ that by quark field in light-front form.
Therefore, we cannot simply drop the boundary integrals for
color singlet states.  In other words, removing the boundary
integrals in eq.(20) implies ignoring the non-vanishing
asymptotic $A_a^i$ fields at longitudinal infinity, and
therefore losses the possibility to address the nontrivial
properties in LFQCD.}}  Clearly, eq.(27) is satisfied only
for physical states.  In perturbation theory, we cannot
use this condition because in perturbative QCD, we consider
not only physical states but also color non-singlet states for
which eq.(27) may not be satisfied.  Therefore
the main effect of eq.(27) should be manifested in
nonperturbative dynamics.  Now we shall discuss how
eq.(27) determines the non-vanishing topological winding number.

To make the discussion clear and without any loss of generality,
we consider the case of zero quark mass only{\cite{zhang2}}.
It is well-known that the axial current for $N_f$-flavor
quarks has an anomalous divergence,
\BE
	\partial_{\mu} j_5^{\mu} = N_f \frac{g^2}{8\pi^2} \tr (F_{\mu \nu}
		\widetilde{F}^{\mu \nu}),
\EE
where the axial current is $j_5^{\mu} = \bar{\psi} \gamma^{\mu} \gamma_5
\psi$, and the dual field strength is $\widetilde{F}^{\mu \nu} = \frac{1}{2}
\epsilon^{\mu \nu \sigma \rho}F_{\sigma \rho}$.  From eq.(28),
the time derivative of the light-front axial charge is given by
\be
	& & \partial^- Q_5 = \int dx^- d^2x_{\bot} \partial^-
		(\psi_+^{\dagger} \gamma_5 \psi_+ ) \nonumber \\
	& & ~~~~~~~~ = N_f \frac{g^2}{8 \pi^2} \int dx^- d^2 x_{\bot} \tr
		(F_{\mu \nu} \widetilde{F}^{\mu \nu}).
\ee
In obtaining eq.(29), we have used the fact that the other
three surface terms of axial currents at longitudinal and
transverse infinity do not contribute to $\partial^- Q_5$.
This is clear if we note that $j_5^i(x) \rightarrow 0$ at $x_{\bot}
\rightarrow \pm \infty$ and $j_5^- (x)= 2 \psi_- (x) \gamma_5
\psi_- (x) $ which leads to $j_5^-|_{x^-=\infty} - j_5^-|_{x^-=- \infty}
=0$ due to the antisymmetric boundary behavior of $\psi_-$ at
longitudinal infinity (see eq.(19)).

The anomaly alone does not imply that the right-hand side
(r.h.s.)~of eq.(29) must be nonzero{\cite{monopole}}.
The nonzero contribution of the r.h.s of eq.(29)
is given by the gauge field configurations determined
by eq.(27).  This can be seen from {\em the winding
number in LFQCD defined as the net charge between
$x^+= - \infty$ and $x^+ = \infty$},
\BE
	\Delta Q_5 = \frac{1}{2} \int dx^+ \partial^- Q_5  = N_f
		\frac{g^2}{8 \pi^2} \int_M d^4x \tr (F_{\mu \nu}
		\widetilde{F}^{\mu \nu}).
\EE
The integration on the r.h.s.~of the above equation is
defined in Minkowski space.  By using the identity
\BE
	\tr (F_{\mu \nu} \widetilde{F}^{\mu \nu}) = 4 \partial_{\mu} K^{\mu}
\EE
where
\BE	K^{\mu} = \epsilon^{\mu \nu \sigma \rho} \tr \left\{
		A_{\nu} \partial_{\sigma} A_{\rho} + \frac{2}{3}
		A_{\nu} A_{\sigma} A_{\rho} \right\} ,
\EE
the r.h.s.~of eq.(30) is reduced to surface integrals.  In the
light-front gauge $A_a^+=0$, the second term in $K^{+,i}$ is zero.
Thus, the surface integrals at transverse infinity vanish
because $K^i$ falls off as $r^{-3}$ for $r = |x_{\bot}| \rightarrow \infty$.
Meanwhile, since $x^+$ and $x^-$ are symmetric, it is reasonable
to use the same boundary condition (i.e., antisymmetric boundary
condition) for $A_a^i$ at $x^+ = \pm \infty$.  Therefore, the
contribution from the surface integral at $x^+$-infinity vanishes
as well.  At $x^-=\pm \infty$, the surface integral contribution
for the first term of $K^-$ is also zero due to the antisymmetric
boundary condition for the $A_a^i$ and $A_a^-$ fields.
Finally, eq.(30) is reduced to{\footnote{As we have seen here,
this solution is obtained by use of eq.(28).  It would be very interesting
if one could directly derive $\Delta Q_5$ or eq.(29) from the LFQCD
Hamiltonian eq.(20).}}
\be
	& & \Delta Q_5 = -N_f \frac{g^2}{ \pi^2} \int dx^+ d^2x_{\bot}
		\tr \left. (A^- [A^1, A^2]) \right|_{x^-=-\infty}^{x^-
		=\infty} \nonumber \\
	& & ~~~~~~~~ = -2N_f \frac{g^2}{ \pi^2} \int dx^+ d^2x_{\bot}
		\tr (A^- [A^1, A^2]), ~~~{\rm at}~ x^-=\infty,
\ee
Here we have used again the {\em antisymmetric} boundary
condition of $A_a^i$ and $A_a^-$ at longitudinal infinity.
Eq.(33) shows that a non-vanishing $\Delta Q_5$ is generated
from the asymptotic fields of $A_a^i, A_a^-$ at longitudinal
infinity, which are generated by the boundary integrals by eq.(27).
The non-zero $A_a^-|_{x^-=\pm \infty}$ is
determined by transverse gauge fields (from Eq.(19)),
\BE
	A_a^-(x^+, x_{\bot})|_{x^-=\pm \infty} = \pm \frac{1}{2}
		\int_{-\infty}^{\infty} dx^- x^- G_a(x^+, x^-, x_{\bot}).
\EE
Thus, the boundary integrals are essential for the non-vanishing
winding number in QCD.

The above derivation shows that
for an antisymmetric boundary condition, although the LFQCD
vacuum is trivial, the nontrivial QCD structure
is switched to the field operators.  This structure is
manifested in the asymptotic behavior of transverse gauge fields
at longitudinal infinity and is explicitly associated with the
boundary integrals.  The trivial vacuum with nontrivial
field variables in the present formulation of LFQCD may provide
a practically useful framework for describing hadronic states.
We now turn to this discussion.

3. {\em Non-local potentials in the transverse direction}.
One of the nonperturbative approaches to solve bound states in
LFFT is the Tamm-Dancoff approach, which truncates the Fock space to
be a few-body state space {\cite{tamm}}. Such an approach becomes
practically applicable only when the vacuum is trivial.  We have
given a realization of a trivial vacuum in this paper.
However, to address hadronic bound states, the existence
of nontrivial potentials, namely, confinement potentials,
is crucial.  An explicit construction of such potentials
from QCD is still lacking. The LFQCD
Hamiltonian contains linear potentials only in the longitudinal
direction (see eq.(20)).  Quark and color confinement certainly
requires similar potentials in the transverse direction as well.
We suggest that these nontrivial potentials in LFQCD might hide
in the condition of eq.(27).

{}From eq.(27) we see that the asymptotic $A_a^i$ fields
at longitudinal infinity are proportional to the color
charge density in transverse space and also that they
involve non-local behavior in the transverse direction (induced by the
transverse derivative).  Intuitively, we may separate the
transverse gauge potentials into a normal part plus a boundary part,
\BE
	A_a^i = A_{aN}^i + A_{aB}^i
\EE
where
\be
	& & A_{aN}^i |_{x^- = \pm \infty}  =0~~,~~
		\partial^i A_{aN}^i |_{x^- = \pm \infty}  = 0 ,\\
	& & \partial^i A_{aB}^i |_{x^- = \pm \infty}  =
		\mp \frac{g}{4}(\rho_a^g(x_{\bot}) + \rho_a^q(x_{\bot})).
\ee
In eq.(38), $\rho_a(x_{\bot})$ denote the color charge densities
integrated over $x^-$, where the definition of the color charge
densities is given in footnote 2.  The conditions of eqs.(36) and
(37) do not uniquely determine the separation of eq.(35).
Generally, there are two types
of separation for eq.(35).  One is to consider $A_{aB}^i$
the long-distance fields generated by the boundary integrals and
$A_{aN}^i$ the short-distance fields determined by free theory.
Thus, $A_{aB}^i$ correspond to the gauge field configuration
for the non-vanishing winding number discussed in the previous subsection.
In this case, if we are only interested in the low-energy dynamics,
the effect of the $A_{aN}^i$ fields may be ignored. Thus, it is very
attractive but is also very difficult to analytically find the
$A_{aB}^i$.  Another possibility is to choose a simple solution
for the $A_{aB}^i$ that satisfy eq.(37).  In this case, the $A_{aN}^i$
have the trivial boundary condition eq.(36) but are not
determined by free theory.  The Hamiltonian is then expressed only
in terms of the $A_{aN}^i$, and the boundary behavior of transverse gauge
fields are replaced by the effective interactions. A convenient
choice for $A_{aB}^i$ which satisfy eq.(37) is
\BE
	A_{aB}^i(x) = - \frac{g}{16} \int dx'^- dx'^i \varepsilon(x^- - x'^-)
		\varepsilon(x^i - x'^i) (\rho_a^g (x') + \rho_a^q(x')) ~,
		~~ i = 1, 2 .
\EE
Substituting the separation of eq.(35) with (38) into the LFQCD Hamiltonian,
we obtain a new Hamiltonian in terms of $A_{aN}^i$ that contains
many effective interactions induced by eq.(27).  All
these effective interactions involve the color charge densities and
involve non-local behavior in both the longitudinal and transverse directions.
One of the lowest order interactions, for example, is given by
\be
	& & H_{b1}
		\propto \sum_{ij} \int_{-\infty}^{\infty} dx^i dx^- dx^j
		dx'^- dx'^j \eta^{ij} \{ \partial^i \rho_a^q(x^-,x^i,x^j)
		|x^- - x'^-| \nonumber \\
	& & ~~~~~~~~~~~~~~~~~~~~~~~~~~~~ |x^j - x'^j| \partial^i
		\rho_a^q(x'^-,x^i,x'^j) \}
\ee
where $\eta^{ij} \equiv 1~ (0) $ for $ i \neq (=)~ j$.
Hence, eq.(37) leads to numerous many-body non-local color
charge interactions which are functions of boundary integrals,
and which may lead to confinement.

As we have mentioned in the introduction, Wilson recently proposed
a formalism to construct a confining light-front quark-gluon
Hamiltonian for LFQCD {\cite{wilsonp}}.  Wilson suggested that
a starting point for
analyzing the full QCD with confinement in light-front form is
the linear infrared divergence (i.e., $1/k^{+2}$ singularity in
momentum space).  Our procedure for constructing effective
interactions from the boundary integrals is indeed
associated with the linear infrared divergences.  However, Wilson's
approach is totally different from what we have discussed here.  His
analysis based on light-front power counting shows that possible confining
potentials may be obtained from the counterterms of the linear infrared
divergence. In QED, the counterterms for infrared divergences are
forbidden because the divergences arise from integration over the
square of the electron scattering amplitude rather than integrals over
the amplitude itself and it is only the latter that can be regulated by
counterterms.  However, in QCD, since quarks and gluons serve only
as constituents, there are no scattering cross section for them, and
there exist counterterms for the linear infrared divergence.  These
counterterms involve the color charge densities integrated over $x^-$
and involve non-local behavior in the transverse direction.  However,
the non-local structure of the counterterms is unknown since
the power counting itself cannot determine it.  They also violate
longitudinal boost invariance so the coefficients of all the
counterterms may further be determined by
the requirement of boost invariance.  Wilson pointed out that
it is tempting to identify these terms as the source of transverse
confinement {\cite{wilsonp}}.

In the present formulation, as we have seen from eq.(27)
or (38), the asymptotic gluon fields which induce effective
interactions are proportional to color charge densities
integrated over $x^-$ (the boundary integrals) and also involve
non-local behavior in the transverse direction. Furthermore,
the non-local behavior of effective interactions in the transverse
direction is also determined by eq.(27) or (38).  Thus, we can
explicitly construct many effective interactions from (38).
It can be shown that the effective interactions related
to fermion part are similar in both QED and QCD but are very
different for the gauge part.  In QED, there is no any effective
interaction generated by the boundary integrals that involves gauge
fields. However, in QCD, there are numerous number of effective
interactions which are coupled to the non-abelian gauge fields.
These effective interactions may be responsible for quark confinement
since they originate from the nontrivial behavior of the non-abelian
transverse gauge fields.  Yet, it is interesting to see
that the analyses based on different approaches have the same
consequence that non-local potentials in the transverse direction
in LFQCD are related to linear infrared divergences which have
not been paid attention in the previous investigations.

Still the Hamiltonian contains, in principle, an infinite number of
many-body interactions generated by the boundary integrals
(or obtained from the counterterms of the linear infrared divergences).
This is a consequence of the boundary integrals in a non-abelian
gauge theory due to the existence of nonlinear gluon interactions.
It is also true in other gauge choices, such as Coulomb gauge
{\cite{chlee}} or axial gauge {\cite{axialg}}.
Practically, as the first step, we may only keep two-body
interaction terms, such as eq.(41), in the new Hamiltonian.
Because of the trivial vacuum in the present formulation of LFQCD,
using such an approximate LFQCD Hamiltonian, we can apply
the light-front Tamm-Dancoff approach to find hadronic bound
states, where the bound states contain only a few particles, such as one
quark-antiquark pair, one quark-antiquark pair with one and two
gluons.  This is certainly one of the most attractive approaches
for low-energy QCD.  A numerical investigation along this consideration
is in progress.

\section{Discussions}
In the previous section, we have discussed some primary properties of
boundary integrals which we think to be important in understanding
LFQCD.  In the current investigations of LFFT, one of the most
active topics is the problem of the $k^+=0$ modes and the
$A^+=0$ gauge.  The implications of the boundary conditions in
determining the nontrivial behavior of gauge theory has,
however, been overlooked.  In this section, we have some remarks
to make about the relation of the $k^+=0$ modes, the $A_a^+=0$
gauge and boundary conditions at longitudinal infinity.

In previous LFFT investigations, much attention has been paid to
how to construct a nontrivial vacuum from the $k^+=0$ modes.
All attempts have focused on $1+1$ field models
{\cite{bmpp}}.  The motivation for these attempts, as
we have mentioned in the introduction, is to try to
understand spontaneous chiral symmetry breaking in LFFT.
In instant form, the vacuum is, of course, crucial for
hadronic structure since we believe that axial charges $Q_5^a$
create pseudoscalar particles (the lowest bound states in strong
interaction region) from the vacuum.  However, the role of light-front
axial charges in hadronic structure is totally different.
The success of light-front current algebra in describing
low-energy hadronic structure is based on the properties of
light-front $Q_5^a$ with a trivial vacuum {\cite{lfca}}.
In this case, $Q_5^a$ annihilate the vacuum so that the
vacuum in LFFT itself is not essential in understanding
chiral symmetry.  The importance of light-front $Q_5^a$
lies in their matrix elements between hadronic states.  These matrix
elements are proportional to hadronic decay constants
involving pseudoscalar mesons and therefore carry the basic
information of hadronic structure.  In instant form, the matrix
elements of $Q_5^a$ in hadronic states with zero momentum
transfer are zero if one does not make use of the infinite
momentum limit {\cite{ff}}.  In other words, the instant
$Q_5^a$ itself is practically not useful for hadronic structure
except for the Nambu-Goldstone picture of spontaneous symmetry
breaking, where the important ingredient is the axial current.
These totally opposite
properties of axial charges in light-front and instant forms
implies that to address dynamical breaking of chiral
symmetry in LFQCD, one may need to understand the relation between
light-front axial charge operators and the Hamiltonian operator
rather than the structure of vacuum in LFFT.  The present
work is motivated by this consideration.

Second, the canonical quantization of light-front gauge theory
is often considered in a box with a periodic boundary condition
for gauge fields.  In this case, it was argued that one cannot
use $A_a^+=0$ for the $k^+=0$ sector because it is incompatible
with the periodic boundary condition in finite volume {\cite{periodic}}.
This argument is true but incomplete.  The $A_a^+=0$ condition does not
totally fix the gauge freedom, and the residual gauge fixing
is responsible for the non-trivial gauge field configurations which,
however, are lacking when one {\em imposes} a periodic boundary
condition [see eq.(33)].  Thus, if one prefers to use a periodic
boundary conditions to quantize LFQCD in a box, one has to choose
other gauges {\cite{periodic}}.  However, for any gauge fixing
other than the $A_a^+=0$, eqs.(9) and (10) show that the constraint
conditions are extremely difficult to solve except for numerical
calculations, such as lattice gauge calculations.  Our treatment in
this paper shows that we can address the non-trivial QCD structure by
use of $A_a^+=0$ gauge and the trivial vacuum if we take into account
the residual gauge fixing in antisymmetric boundary conditions for
field variables at longitudinal infinity. In principle for $A_a^+=0$
gauge, other boundary conditions can also be used {\cite{steinhardt}}
yet the resulting theory is currently intractable due to the existence
of the $k^+=0$ modes.

Finally, we discuss briefly the difference between LFQCD and
the canonical formulation of QCD in instant form with axial
gauge $A_a^3=0$.  The main difference is as follows.  In LFQCD,
the finite energy density for physical states results explicitly
in the asymptotic equation for transverse gauge fields in
longitudinal infinity [see eq.(27)] due to the fact
that the conjugate momenta of $A_a^i$ are dependent variables in
light-front form, ($E_a^i = \frac{1}{2} \partial^+ A_a^i$ is not
a light-front time derivative).  In axial gauge, $A_a^i, i=1,2$ and
their conjugate momenta are all the dynamically independent variables.
Thus, a similar condition proposed by Chodos leads
to a very complicated formalism which may not be practically useful
even for perturbation theory, as noted by himself {\cite{axialg}}.
The second major difference is the vacuum.  In axial gauge, the QCD
vacuum is still complicated regardless of the boundary condition
chosen.  In such a case, it is very difficult (if not impossible)
to do non-perturbative calculations before knowing the vacuum
structure.  In LFQCD, with antisymmetric boundary conditions,
the vacuum is trivial and the nontrivial behavior of QCD would be
manifested directly in Hamiltonian operators induced by the
boundary integrals.  Thus it is straightforward to use quantum mechanical
non-perturbative approaches to compute bound states.  Moreover,
in axial gauge, the boost invariance is not manifested kinematically
so that it is not a good framework  to study low-energy
QCD, which deals with composite particles of quarks and gluons.
In LFQCD, as we have mentioned in the introduction, boost invariance
is a kinematical symmetry which is very useful
in addressing hadronic structures.

In summary, the essential point in determining nontrivial
behavior of LFQCD seems to be the boundary conditions.
A suitable choice of boundary condition for physical fields in
LFQCD is crucial because it determines whether the nontrivial
behavior of QCD can be decoupled from the vacuum so that
the property of the trivial vacuum in LFFT becomes useful
for solving hadrons from QCD.  We have derived the canonical
formulation of LFQCD with great care for boundary integrals,
which have not been paid enough attention in previous
investigations.  We show that the boundary integrals
are the source of the light-front linear infrared singularity
and involve color charge densities and non-local behavior
in the transverse direction that lead to non-local forces
generated by the boundary integrals which are also responsible
for the non-vanishing topological winding number.  Clearly, our
understanding of the physics from the boundary integrals
in LFQCD is far from complete and much work remains to be done.
Particularly, two questions are very interesting for the
understanding of hadronic physics.  One is which terms among the
numerous non-local interactions are essential for hadronic bound
states.  The other is how we can find an explicit
field configuration for the non-zero winding number
in LFQCD that satisfies the asymptotic behavior of eq.(27). These
are two of the main problems in nonperturbative LFQCD for the future.

\section*{Acknowledgement}
We would like to thank R. J. Perry for many critical comments and
helpful discussions during this work.  We would also like to
acknowledge fruitful discussions with R. J. Furnstahl, K. Hornbostel,
J. Shigemitsu, T. S. Walhout, and K. G. Wilson.  We want to specially
thank R. J. Furnstahl for his carefully reading of the manuscript
and many useful comments for the paper.
This work was supported by National Science Foundation of United States
under Grant No. PHY-9102922, PHY-8858250 and PHY-9203145.

\appendix
\renewcommand{\thesection}{Appendix \Alph{section}.}
\renewcommand{\thesubsection}{\Alph{section}-\arabic{subsection}.}
\renewcommand{\theequation}{\Alph{section}.\arabic{equation}}
\section{Canonical quantization of LFQCD}
\setcounter{equation}{0}
A self-consistent formulation of LFQCD requires that the resulting
Hamiltonian must generate the correct equations of motion
for the physical degrees of freedom $(A_a^i, \psi_+, \psi_+^{\dagger})$.
This appendix is devoted to the derivation of canonical quantization
and a check of the consistency.

To see how to correctly reproduce the Lagrangian equations of motion,
we need to find consistent commutators for physical field
variables.

In the light-front gauge, the Lagrangian of eq.(7) is reduced to
\BE
	{\cal L} = \frac{1}{2} \partial^+ A_a^i \partial^- A_a^i
		+ \frac{i}{2} (\psi_+^{\dagger} \partial^- \psi_+ -
		\partial^- \psi_+^{\dagger} \psi_+ ) - {\cal H}.
\EE
The canonical momenta of the physical field variables $A_a^i, \psi_+,
\psi_+^{\dagger}$ are
\BE
	{\cal E}_a^i = \frac{1}{2} \partial^+ A_a^i ~,~~
		\pi_{\psi_+} = \frac{i}{2} \psi_+^{\dagger} ~,~~
		\pi_{\psi_+^{\dagger}} = - \frac{i}{2} \psi_+.
\EE
However, eq.(A.2) shows that all the canonical momenta are functions
of the independent field variables.  Thus, after determining
all the Lagrange multipliers, the system is still a constrained
Hamiltonian system.  Usually, in order to quantize such a constrained
Hamiltonian system, one has to use the Dirac procedure, by imposing
the so-called primary, second-class constraints $E_a^i+ \frac{1}{2}
\partial^+ A_a^i =0$ (similarly for $\pi_{\psi_+, \psi_+^{\dagger}}$)
to construct Dirac brackets.  However, for these trivial primary
constraints, the mathematically well-defined canonical one-form
offers a rigorous phase space structure for canonical
quantization {\cite{fjc}}.{\footnote{If the first-class constraints
cannot be solved explicitly, for example for other gauge choices in
light-front form, using the Dirac procedure may be necessary to
construct the canonical commutation relations for all
variables.}}  In this appendix, we will use such an approach
for light-front quantization.

The phase space structure for the physical variables (${\cal E}_a^i, A_a^i;
\pi_{\psi_+}, \psi_+; \pi_{\psi_+^{\dagger}},\psi_+^{\dagger} $) is determined
rigorously by rewriting eq.(A.1) as a Lagrangian one-form ${\cal L}dx^+$
(apart from a total light-front time derivative),
\be
	& & {\cal L}dx^+ = \frac{1}{2} 2 ({\cal E}_a^i dA_a^i + \pi_{\psi_+}
		d\psi_+ + d\psi_+^{\dagger}\pi_{\psi_+^{\dagger}} - A_a^i
		d{\cal E}_a^i - d\pi_{\psi_+}\psi_+ - \psi_+^{\dagger}
		d\pi_{\psi_+^{\dagger}} ) - {\cal H} dx^+ \nonumber \\
	& & ~~~~~~~~= \frac{1}{2} q^{\alpha} \Gamma_{\alpha \beta}
		dq^{\beta} - {\cal H} dx^+
\ee
where the first term in the right-hand side is called the canonical
one-form of the physical phase space, and quark fields are anticommuting
{\em c}-numbers (Grassmann variables).  Correspondingly, the symplectic
structure or the Poisson brackets of the phase space is given by
\BE
	\omega = \frac{1}{2} \Gamma_{\alpha \beta} dq^{\alpha} dq^{\beta} ~~
		{\rm or} ~~ [q^{\beta} , q^{\alpha}]_p = \Gamma_{\alpha
		\beta}^{-1} .
\EE
Canonical quantization is realized by replacing the Poisson brackets
by the equal-$x^+$ commutation relations
\BE
	[q^{\beta},  q^{\alpha}] = i \Gamma_{\alpha \beta}^{-1} .
\EE
Explicitly
\be
	& & [A_a^i (x) , {\cal E}_b^j(y) ]_{x^+=y^+} = i \frac{1}{2}
		\delta_{ab} \delta^{ij} \delta^3 (x-y), \\
	& & \{\psi_+ (x), \pi_{\psi_+}(y) \}_{x^+=y^+} = i \frac{\Lambda_+}
		{2} \delta^3 (x-y) , \\
	& & \{\psi_+^{\dagger} (x), \pi_{\psi_+^{\dagger}}(y) \}_{
		x^+=y^+} = -i \frac{\Lambda_+}{2} \delta^3 (x-y)
\ee
or
\be
	& & [A_a^i (x) , \partial^+ A_b^j(y) ]_{x^+=y^+} = i \delta_{ab}
		\delta^{ij} \delta^3 (x-y) , \\
	& & [A_a^i (x) , A_b^j(y) ]_{x^+=y^+} = - i \delta_{ab} \delta^{ij}
		\frac{1}{4} \varepsilon(x^- - y^-) \delta^2(x_{\bot}-
		y_{\bot}) , \\
	& & \{\psi_+ (x), \psi_+^{\dagger}(y) \}_{x^+=y^+} =
		\Lambda_+ \delta^3(x-y)
\ee
where $\delta^3(x-y) \equiv \delta(x^--y^-) \delta^2 (x_{\bot} - y_{\bot})$.
All other commutators between the physical degrees of freedom vanish.
Note that, unlike in the instant form, the commutator $[A_a^i (x),
A_a^j(y) ]$ does not vanish since the canonical momentum is a function of
the coordinates.  Eq.(A.10) is defined consistently with the antisymmetric
boundary condition, $A_a^i(-\infty) = - A_a^i(\infty)$, where it is also
required that
\BE
	\lim_{x^- \rightarrow \infty, y^- \rightarrow -\infty}
		\varepsilon (x^- - y^-) = 0
\EE
Topologically, this requirement is identical to $\varepsilon(0) = 0 $.
This can easily be understood if we divide the longitudinal line
into boxes and define the $\varepsilon$-function in each box.

Using the above basic commutation relations, it is straightforward
to verify that the equations of motion are consistent with eqs.(2)
and (3),
\be
	& & \partial^- \psi_+ = \frac{1}{i} [ \psi_+ ~,~ H] \nonumber \\
	& & ~~~~~~~~ = \left\{ i g A^- - \frac{1}{4} \{ \alpha_{\bot} \cdot
		( i \partial_{\bot} + g A_{\bot}) + \beta m \} \right.
		\nonumber \\
	& & ~~~~~~~~~~~~~~~~~~~ \times \left. \int_{-\infty}^{\infty} dx'^-
		\varepsilon (x^- - x'^-) \{ \alpha_{\bot} \cdot ( i
		\partial_{\bot} + g A_{\bot}) + \beta m \} \right\} \psi_+ \\
	& & \partial^- A_a^i = \frac{1}{i} [ A_a^i ~,~ H] \nonumber \\
	& & ~~~~~~~~ = \frac{1}{4} \int_{-\infty}^{\infty} dx'^- \varepsilon
		(x^- - x'^-) [ D_{ab}^j F_b^{j i}(x^+,x'^-,x_{\bot}))
		- D_{ab}^i E_b^-(x^+,x'^-,x_{\bot}) \nonumber \\
	& & ~~~~~~~~~~~~~~~~~~~~ - g j_a^i(x^+,x'^-,x_{\bot})
		- g f^{abc} (A_b^- \partial^+ A_c^i)(x^+,x'^-,x_{\bot})]
\ee
where $D_{ab}^i = \delta_{ab} \partial^i - g f^{abc}A_c^i$, and $A_a^-$ and
$E_a^-$ are given by eqs.(19) and (15).

\section{Cancellation of linear infrared divergence}
\setcounter{equation}{0}
In this appendix, we shall use the $x^+$-ordering perturbative rule
which we developed recently{\cite{zhang1}} for the two-component
LFQCD to check the cancellation of linear infrared divergence in
perturbative LFQCD.  For the reader's convenience, we list some
relevant diagrammatic rules in Table 1 for the following
calculation. For a complete list of the $x^+$-ordering perturbative
rules and Feynman rules for two-component LFQCD, see Ref.{\cite{zhang1}}.

\subsection{Tree level ($q\bar{q}$ scattering)}
The lowest-order $q\bar{q}$ scattering amplitude is given by
\BE
	M_{fi} = M_{fi}^a + M_{fi}^b + M_{fi}^c ,
\EE
which corresponds to the diagrams shown in Fig.1.  Using the rules listed
above and the $x^+$-ordering perturbative theory, we immediately obtain that
\be
	& & M_{fi}^a + M_{fi}^b = g^2 T_{21}^a T_{43}^a \chi_2^{\dagger}
		\Gamma_0^q(p_2,p_1) \chi_1 \chi_4^{\dagger} \Gamma_0^q
		(p_4,p_3) \chi_3 \nonumber \\
	& & ~~~~~~~~~~~~~~~~~ \left\{ \frac{1}{p_i^- - p_1^- - p_4^- + k^-}
		+ \frac{1}{p_i^- - p_2^- - p_3^- - k^-}
		\right\}, \\
	& & M_{fi}^c = \left\{ \begin{array}{l}	2g^2 T_{21}^a T_{43}^a
		\chi_2^{\dagger} \chi_1 \frac{1}{[k^{+2}]}
		\chi_4^{\dagger} \chi_3 ~~~~~~ 	{\rm NB}, \\
		2g^2 T_{21}^a T_{43}^a \chi_2^{\dagger} \chi_1 \frac{1}{[k^+]^2}
		\chi_4^{\dagger} \chi_3 ~~~~~~~~ 	{\rm WB}
		    \end{array} \right.
\ee
where
\BE
	\Gamma_0^q(p_2,p_1) = \left[2\frac{k^i}{[k^+]}
		- \frac{\sigma \cdot p_{2\bot}-im}{[p_2^+]} \sigma^i -
		\sigma^i \frac{\sigma \cdot p_{1\bot} + im}{[p_1^+]} \right],
\EE
$p_i^-$ is the total energy of the initial state, $k^{\mu} = (p_1^+ - p_2^+,
p_1^i - p_2^i, \frac{(p_{1\bot} - p_{2\bot})^2 +m^2}{p_1^+ - p_2^+})$.
NB denotes no boundary term and WB means including the boundary
term.  It follows that in the principal value prescription,
$M_{fi}^a + M_{fi}^b$ is free of infrared divergences, while $M_{fi}^c$
without the boundary term has a $1/\epsilon^2$ divergence when $k^+
\rightarrow 0$.  When the boundary term is included, the $1/\epsilon^2$
is cancelled [see (B.3)].  Therefore, it is necessary to include the
boundary term in order to obtain a finite amplitude for the lowest-order
$q \bar{q}$ scattering.  A similar discussion for $e^+ e^-$ scattering
in LFQED is given in ref.{\cite{Tang}}.

\subsection{Loop corrections (one-loop quark self-energy)}
Based on the $x^+$-ordering perturbative theory, the quark on-shell
self-energy (mass correction) up to one-loop is determined by
\BE
	\Sigma(p^2=m^2) = \Sigma_a + \Sigma_b + \Sigma_c.
\EE
The three terms in the right-hand side are denoted by the three
diagrams shown in Fig.2.  Again, using the rules listed above,
we find that
\be
	& & \Sigma_a = g^2 C_f \int \frac{dk^+ d^2k_{\bot}}{16\pi^3}
		\frac{\theta(k^+) \theta(p^+-k^+)}{[k^+]} \left\{
		2\frac{k^i}{[k^+]} - \frac{\sigma \cdot p_{\bot}-im}{[p^+]}
		\sigma^i \right.\nonumber \\
	& & ~~~~~~~~~~~~~~~~~~ \left. - \sigma^i \frac{\sigma \cdot p_{\bot}
		-\sigma \cdot k_{\bot} + im}{[p^+ - k^+]} \right\} \left\{
		2\frac{k^i}{[k^+]} - \frac{\sigma \cdot p_{\bot} - \sigma
		\cdot k_{\bot} -im}{[p^+ - k^+]} \sigma^i \right. \nonumber \\
	& & ~~~~~~~~~~~~~~~~~~ \left. - \sigma^i \frac{\sigma \cdot p_{\bot}
		+ im }{[p^+]} \right\} \frac{1}{p^- - k^- - (p-k)^-} \\
	& & \Sigma_b = 2g^2 C_f \int \frac{dk^+ d^2k_{\bot}}{16\pi^2}
			\frac{1}{[k^+][p^+-k^+]} \\
	& & \Sigma_c = \left\{ \begin{array}{l} 2g^2 C_f \int \frac{dk^+
		d^2k_{\bot}}{16\pi^2}
		\left\{ \frac{1}{[(p^+-k^+)^2]} - \frac{1}{[(p^+ +
		k^+)^2]}  \right\} ~~~~~~ {\rm NB} \\
		2g^2 C_f \int \frac{dk^+ d^2k_{\bot}}{16\pi^3}
		\left\{	\frac{1}{[p^+-k^+]^2} - \frac{1}{[p^+ + k^+]^2}
		\right\} ~~~~~~~~~~ {\rm WB}
		\end{array} \right.
\ee
where $C_f= 4/3 $.  A direct calculation shows that
\be
	& & \Sigma_a = \frac{g^2}{8\pi^2} \frac{C_f}{p^+} \int
		d^2k_{\bot} \left( \int_0^1 dx \frac{2m^2}{k_{\bot}^2
		+ x^2m^2} + 1 - \frac{\pi p^+}{2\epsilon} + \ln
		\frac{\epsilon}{p^+} \right) \\
	& & \Sigma_b = -\frac{g^2}{8\pi^2} \frac{C_f}{p^+} \int d^2k_{\bot}
		\left( \ln \frac{\epsilon}{p^+} \right) \\
	& & \Sigma_c = \left\{ \begin{array}{l}
		\frac{g^2}{8\pi^2} \frac{C_f}{p^+} \int d^2k_{\bot}(-2)
		~~~~~~~~~~~~~~~~~~~~~ {\rm NB} \\
		\frac{g^2}{8\pi^2} \frac{C_f}{p^+} \int d^2k_{\bot}
		\left(-1 + \frac{\pi p^+}{2\epsilon} \right)~~~~~~~~~~
		{\rm WB} \end{array} \right.
\ee
which tells us that in the one-loop correction to the quark self-energy,
the one-gluon exchange contains both linear and logarithmic
infrared divergences.  The instantaneous fermion interaction
contains only one logarithmic divergence (see $\Sigma_b$
in eq.(B.10)), which cancels the logarithmic divergence in
$\Sigma_a$.  The naive instantaneous-gluon interaction
(namely the linear potential in the longitudinal direction)
is free of infrared divergence.  Therefore, without boundary term,
the quark mass correction involves a linear infrared divergence,
which is an inconsistent solution.  By combining the boundary
term with the linear potential, we see that $\Sigma_c$
has a linear infrared divergence which precisely cancels the
same divergence in $\Sigma_a$.  Thus the quark mass correction is
now free of infrared divergences,
\BE
	\delta m^2 = p^+ \Sigma = \frac{m^2}{4\pi^2} C_f \ln
		\frac{\Lambda^2}{m^2} + {\rm finite}
\EE
where $\Lambda$ is the transverse momentum cut-off.  In eq.(B.9), the
coefficient (1/4) in the mass correction is different from the covariant
result (3/8) because the regularization scheme is different.  This
coefficient is the same as that in the light-front calculation with dimensional
regularization in the transverse direction and the explicit cutoff in
the longitudinal direction {\cite{lfqedh}}.
Note that in their calculation,
the expressions of eqs.(B.9--11) are different but the sum is the same as
eq.(B.12) where the linear divergence is also cancelled.


\newpage
\input FEYNMAN
\THICKLINES

{\Large\bf Table 1:} Some $x^+$-ordering diagrammatic rules for the
two-component LFQCD:

\begin{picture}(38000,9000)
\drawline\fermion[\E\REG](1000,3000)[9000]
\drawarrow[\LDIR\ATTIP](3000,3000)
\drawarrow[\LDIR\ATTIP](9000,3000)
\put(2500,2000){$p_1$}
\put(8500,2000){$p_2$}
\put(0,3000){$\alpha$}
\put(10500,3000){$\beta$}
\drawline\gluon[\NE\REG](5500,3000)[3]
\put(\pmidx,6500){$k$}
\put(10000,7000) {$i$}
\put(15000,4000) {\( - g (T^a) \chi_{\beta}^{\dagger} \left\{ 2
		\frac{k^i}{[k^+]} - \frac{ \sigma \cdot p_{2\bot} -
		im}{[p_2^+]} \sigma^i - \sigma^i \frac{ \sigma \cdot
		p_{1\bot} + im}{[p_1^+]} \right\} \chi_{\alpha}
		\epsilon^{i*} \) }
\end{picture}

\begin{picture}(38000,10000)
\drawline\fermion[\E\REG](5500,3000)[4500]
\drawarrow[\LDIR\ATTIP](9000,3000)
\put(8500,2000){$p_2$}
\drawline\fermion[\E\REG](1000,7400)[4500]
\drawarrow[\LDIR\ATTIP](3000,7400)
\put(2500,8400){$p_1$}
\put(0,7400){$\alpha$}
\put(10500,3000){$\beta$}
\drawline\fermion[\N\REG](5500,3000)[4400]
\drawline\gluon[\E\REG](5500,7400)[4]
\put(8500,8400){$k_1$}
\drawline\gluon[\W\REG](5500,3000)[4]
\put(2500,2000){$k_2$}
\put(0,3000){$j$}
\put(10500,7400){$i$}
\drawline\fermion[\E\REG](5100,5200)[800]
\put(15000,4000){$ g^2 (T^a T^b) \chi_{\beta}^{\dagger}
	\frac{\sigma^i \sigma^j}{[p_1^+ + k_1^+]} \chi_{\alpha}
	\epsilon^{i*} \epsilon^j $}
\end{picture}

\begin{picture}(38000,10000)
\drawline\fermion[\E\REG](1000,3000)[9000]
\drawarrow[\LDIR\ATTIP](3250,3000)
\drawarrow[\LDIR\ATTIP](8750,3000)
\put(2700,2000){$p_3$}
\put(7800,2000){$p_4$}
\put(0,3000){$\gamma$}
\put(10500,3000){$\delta$}
\drawline\fermion[\E\REG](1000,7400)[9000]
\drawarrow[\LDIR\ATTIP](3250,7400)
\drawarrow[\LDIR\ATTIP](8750,7400)
\put(2700,8200){$p_1$}
\put(7800,8200){$p_2$}
\put(0,7400){$\alpha$}
\put(10500,7400){$\beta$}
\drawline\gluon[\N\REG](5500,3000)[4]
\drawline\fermion[\E\REG](5000,5200)[1000]
\put(15000,5200){$ \left\{ \begin{array}{l}
	2g^2 (T^a T^a) \chi_{\beta}^{\dagger}
		\chi_{\alpha} \frac{1}{[(p_1^+ - p_2^+)^2]}
		\chi_{\delta}^{\dagger} \chi_{\gamma}  ~~~~~~ {\rm NB} \\
	2g^2 (T^a T^a) \chi_{\beta}^{\dagger}
		\chi_{\alpha} \frac{1}{[p_1^+ - p_2^+]^2}
		\chi_{\delta}^{\dagger} \chi_{\gamma} ~~~~~~~ {\rm WB}
		\end{array} \right.  $}
\end{picture}

Remarks.  The $\chi_{\alpha}$ are two-component
light-front quark spinors, $\chi_{\uparrow} =\left( _0^1 \right)$ and
$\chi_{\downarrow}= \left( _1^0 \right)$, and the $\epsilon^i$ are
two-component gluon polarization vectors.
We have also used the Majorana representation of the $\gamma$ matrices
in the above realization of the two-component formulation.
In additional, the internal integral for quark momentum
is given by $\int \frac{dp^+ d^2p_{\bot}}{16\pi^2} \theta(p^+)$, while
for gluon it is $\int \frac{dk^+ d^2k_{\bot}}{16\pi^2} \frac{\theta(k^+)}
{[k^+]}$.  The light-front free quark and gluon energies are $p^- =
\frac{p_{\bot}^2 + m^2}{[p^+]}$ and $k^- = \frac{k_{\bot}^2}{[k^+]}$

\newpage

{\Large\bf Figures:}

\begin{picture}(38000,15000)
\drawline\fermion[\E\REG](1000,3150)[9000]
\drawarrow[\W\ATBASE](2750,3150)
\drawarrow[\W\ATBASE](8250,3150)
\put(2700,2150){$p_3$}
\put(7800,2150){$p_4$}
\drawline\fermion[\E\REG](1000,7000)[9000]
\drawarrow[\LDIR\ATTIP](3250,7000)
\drawarrow[\LDIR\ATTIP](8750,7000)
\put(2700,7800){$p_1$}
\put(7800,7800){$p_2$}
\drawline\gluon[\NW\FLIPPED](7000,3150)[3]
\put(5000,0){ \( (a) \) }
\put(11500,5100){ $ + $ }
\drawline\fermion[\E\REG](15000,3150)[9000]
\drawarrow[\W\ATBASE](16750,3150)
\drawarrow[\W\ATBASE](22250,3150)
\put(16700,2150){$p_3$}
\put(21800,2150){$p_4$}
\drawline\fermion[\E\REG](15000,7000)[9000]
\drawarrow[\LDIR\ATTIP](17250,7000)
\drawarrow[\LDIR\ATTIP](22750,7000)
\put(16700,7800){$p_1$}
\put(21800,7800){$p_2$}
\drawline\gluon[\NE\REG](18000,3150)[3]
\put(19000,0){ \( (b) \) }
\put(25500,5100){ $ + $ }
\drawline\fermion[\E\REG](29000,3000)[9000]
\drawarrow[\W\ATBASE](30750,3000)
\drawarrow[\W\ATBASE](36250,3000)
\put(30700,2000){$p_3$}
\put(35800,2000){$p_4$}
\drawline\fermion[\E\REG](29000,7400)[9000]
\drawarrow[\LDIR\ATTIP](31250,7400)
\drawarrow[\LDIR\ATTIP](36750,7400)
\put(30700,8200){$p_1$}
\put(35800,8200){$p_2$}
\drawline\gluon[\N\REG](33500,3000)[4]
\drawline\fermion[\E\REG](33000,5200)[1000]
\put(33000,0){ \( (c) \) }
\end{picture}

\(\ \\ \)

\begin{description}
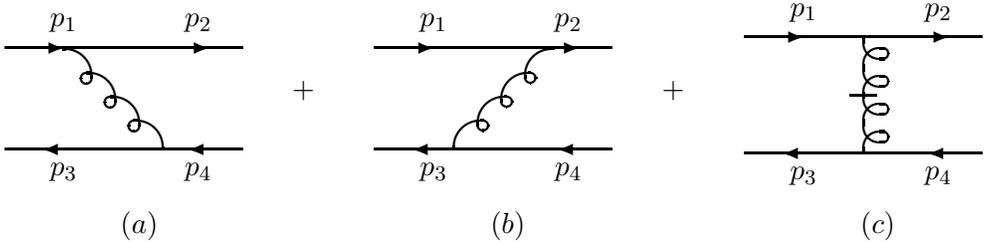
\item [{\bf Fig.1.}] The $x^+$-ordering graphs for the lowest-order $q\bar{q}$
scattering in perturbative LFQCD.
\end{description}

\begin{picture}(38000,15000)
\drawline\fermion[\E\REG](1000,3000)[9000]
\drawarrow[\LDIR\ATTIP](2750,3000)
\drawarrow[\LDIR\ATTIP](9250,3000)
\drawarrow[\LDIR\ATTIP](6000,3000)
\put(2200,2000){$p$}
\put(8500,2000){$p$}
\put(5500,6300){$k$}
\drawloop\gluon[\N 5](3150,3000)
\put(5000,0) {\( (a) \)}
\put(12000,4500) {$+$}
\drawline\fermion[\E\REG](19500,2700)[4500]
\drawarrow[\LDIR\ATTIP](22950,2700)
\put(22200,1700){$p$}
\drawline\fermion[\E\REG](15000,7600)[4500]
\drawarrow[\LDIR\ATTIP](17050,7600)
\put(16500,6600){$p$}
\drawline\fermion[\N\REG](19500,2700)[4900]
\drawline\fermion[\E\REG](19100,5150)[800]
\drawloop\gluon[\SE 3](19500,7600)
\put(22000,5000){$k$}
\put(19000,0) {\( (b) \)}
\put(26000,4500) {$+$}
\drawline\fermion[\E\REG](32500,3000)[5000]
\drawarrow[\LDIR\ATTIP](36250,3000)
\put(35500,2000){$p$}
\drawline\fermion[\E\REG](29000,7400)[3500]
\drawarrow[\LDIR\ATTIP](30750,7400)
\put(30200,6400){$p$}
\put(31300,4800){$k$}
\drawline\gluon[\N\REG](32500,3000)[4]
\put(32500,5200){\oval(4400,4400)[r]}
\drawline\fermion[\E\REG](32000,5200)[1000]
\put(32000,0) {\( (c) \)}
\end{picture}
\(\ \\ \)
\begin{description}
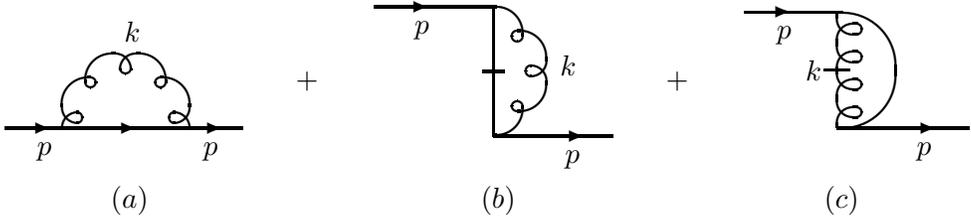
\item [{\bf Fig.2.}] The $x^+$-ordering graphs for the one-loop correction of
quark self-energy in perturbative LFQCD.
\end{description}

\end{document}